\documentclass[review,11pt]{elsarticle}

\usepackage{lineno,hyperref}
\usepackage{amsmath}
\usepackage{color}

\usepackage{graphicx}
\graphicspath{ {images/} }
\usepackage{bm}
\usepackage[export]{adjustbox}

\makeatletter
\def\ps@pprintTitle{%
 \let\@oddhead\@empty
 \let\@evenhead\@empty
 \def\@oddfoot{}%
 \let\@evenfoot\@oddfoot}
\makeatother

\begin{document}

\begin{frontmatter}

\title{Observation of adiabatic and non-adiabatic behavior for CPMG sequence in time-dependent magnetic fields}

\author[SDR]{Martin D. H{\"u}rlimann\corref{mycorrespondingauthor}}
\ead{mhurlimann@gmail.com}

\author[SDR]{Shin Utsuzawa}
\ead{SUtsuzawa@slb.com}

\author[SDR]{Chang-Yu Hou}
\ead{CHou2@slb.com}

\cortext[mycorrespondingauthor]{Corresponding Author}
\address[SDR]{Schlumberger-Doll Research, Cambridge, MA 02139, USA}

\begin{abstract}

We investigate experimentally  the effect of time dependent magnetic fields on the spin dynamics of the Carr - Purcell - Meiboom - Gill (CPMG) sequence. Over a wide range of offset fields and ramp rates, the measured response is fully consistent with adiabatic behavior. The echo amplitudes exhibit characteristic modulations that are in excellent agreement with theoretical predictions. Non-adiabatic events are observed at distinct offsets. Abruptly after passing through these offsets, the experimental results  deviate from the theoretical adiabatic expressions. These non-adiabatic events occur precisely at the field offsets predicted by theory.  It is demonstrated that in the adiabatic regime the effects of field fluctuations are fully reversible, while the occurrence of non-adiabatic events leads to hysteresis. The adiabatic range of field offsets can be increased by modifying the refocusing pulses within the CPMG sequence.

%
%
%

\end{abstract}


\end{frontmatter}

\section{Introduction}

The Carr-Purcell-Meiboom-Gill (CPMG) sequence \cite{Carr1954,Meiboom1958} has found wide-spread application for a series of purposes that include the  monitoring of dynamic processes \cite{stewart1993,loria1999}, the enhancement of signal \cite{lim2002}, and the reduction of decoherence from external noise \cite{viola98,Cappellaro2006,biercuk2009}. This sequence also underlies essentially all applications in  grossly inhomogeneous fields \cite{Hurlimann2000,hurlimann_nmr_2015}. In quantum measurement protocols the technique is usually referred to as 'dynamic decoupling' \cite{viola98}. 

The sequence has been designed to be implemented in a static external magnetic field. Here we consider the case when the external field becomes time-dependent with an amplitude of fluctuation that can become comparable or larger than the strength of the RF field. We demonstrate experimentally that the response can be characterized by adiabatic behavior that is interrupted by non-adiabatic events. Understanding this problem is of practical importance for the interpretation of experiments in unstable fields or when there is relative motion between the sample and measurement apparatus. We recently presented a theoretical treatment based on an eigenmode analysis of the echo propagator \cite{Hurlimann2019}. This is reviewed in section \ref{sec:theory}. Section \ref{sec:setup} presents the experimental set-up and section \ref{sec:experiments} presents the measurements and main results. The conclusion in section \ref{sec:conclusions} discusses the implication of these results. 

\section{Theoretical considerations}
\label{sec:theory}

The CPMG experiment consist of an initial excitation pulse followed by a long string of refocusing pulses, separated by the echo spacing $t_E$. The magnetization is probed stroboscopically at the nominal echo times $k t_E$, where $k$ is a positive integer. 
The theoretical analysis in \cite{Hurlimann2019} is based on the inspection of a single refocusing cycle. The echo-to-echo propagator of the magnetization is decomposed into its eigenmodes. Critically, one of the eigenmodes referred to as the 'CPMG mode' has unity eigenvalue. The corresponding eigenvector ${\hat n}$ points along the  effective field ${\vec B}_{eff}$ that characterizes the average Hamiltonian of a refocusing cycle. The initial excitation pulse of the CPMG pulse sequence is designed to populate this particular eigenmode. Magnetization of this eigenmode is consistently refocused from echo to echo, even in the presence of field inhomogeneities. The evolution of this mode is characterized by the absence of a dynamic and geometric phase. In contrast, the other eigenmodes (usually referred to as CP modes) have eigenvalues of the form $e^{\pm i\alpha}$. The phase $\alpha$ depends on the offset between Larmor and RF frequency, $\omega_0 \equiv \gamma B_0 - \omega_{RF}$, as shown explicitly in \ref{app:A}. 
In the presence of spatial field inhomogeneities across the sample, this dispersion gives rise to a rapid echo to echo dephasing of the CP magnetization component, in analogy to common $T_2^*$ dephasing of the free induction decay. Therefore, after the first few refocusing cycles, only the CPMG component typically contributes to the detected signal. Explicit expressions for the eigenvectors and eigenvalues, including their dependence  on $\omega_0$ are given in the appendix.

When the external field is changed sufficiently slowly, the magnetization follows the evolving eigenmodes adiabatically. The magnetization of the CPMG components is effectively spin-locked to its eigenvector, as shown in Eq.\,\ref{eq:S_fullyadiabatic} in \ref{app:B}. On resonance, this eigenvector lies in the transverse plane, but it acquires a longitudinal component as the amplitude of the external field is increased. This change in direction is a non-monotonic function of $\omega_0$ and results in a systematic modulation of the detected signal as the applied field is changed. In the adiabatic regime, the change of signal induced by a fluctuating external field is reversible and independent of the path of the fluctuation.

To stay in this adiabatic regime, the adiabatic condition ${\cal A} \gg 1$ has to be fulfilled. As discussed in detail in \cite{Hurlimann2019}, the adiabaticity parameter ${\cal A}\equiv \nu_{0,crit} / (d {\tilde \omega}_0 /d\tau)$ is the ratio of the instantaneous critical velocity $\nu_{0,crit}$, an intrinsic parameter, and the dimensionless ramp rate of the external field, 
$d {\tilde \omega}_0 /d\tau \equiv d \frac{\omega_0(t)}{\omega_1} / d \frac{t}{t_E}$, an experimental parameter. The critical velocity $\nu_{0,crit}$ is controlled by the change of the direction $\theta$ of the eigenvector $\hat n$ with offset frequency $\omega_0$ and the strength of the effective field $\alpha = \gamma |{\vec B}_{eff}|$: $ \nu_{0,crit} \equiv {\alpha}/({{d\theta}/{d {\tilde \omega}_0}})$. 

At distinct offset frequencies where the modes become nearly degenerate, the critical velocity has pronounced minima. When the variable external field reaches a value corresponding to one of these minima, the spin dynamics will likely become non-adiabatic, except when the ramp rates becomes very small. Transitions between the CPMG and CP modes then occur. After such a non-adiabatic event,  any generated CP magnetization component will quickly dephase in the presence of any small field inhomogeneities.  The detected signal is controlled mainly by the remaining occupation of the CPMG mode.

\section{Experimental Set-up}
\label{sec:setup}

Experiments were performed on a cylindrically shaped water sample (11 mm diameter, 10 mm length) that was inserted in a solenoid rf coil and placed in a horizontal superconducting  imaging magnet (Nalorac) with a 30 cm bore. The rf frequency was set to 85.1 MHz, the Larmor frequency of the magnet. The rf power was adjusted so that the duration of the nominal $180^\circ $ pulse resulted in  $t_{180}  = 40 \,\mu$s. This corresponds to a nutation frequency $\frac{\omega_1 }{2\pi} = \frac{1}{2 t_{180}} = $ 12.5 kHz. 

The $B_0$ field applied to the sample was varied by up to 1.29 mT, corresponding to a change in Larmor frequency of $4.4\,\omega_1$. This was accomplished using a commercial gradient set (Bruker BGA12SL) that was physically offset by 8 cm along its axis. At the site of the sample, this arrangement generated an essentially uniform bias field (rather than a gradient field in the center of the gradient set). The strength and time dependence of this bias field was controlled through the spectrometer by adjusting the current through the gradient set-up. The spatial uniformity of the bias field across the sample can be approximated by a Gaussian distribution with a relative width of $6.7\times 10^{-3}$. 
The bias field was typically ramped up linearly  from zero to a maximum value with a constant ramp rate. Measurements were also performed with a linear ramp-up immediately followed by a linear ramp-down.  We present results for normalized ramp rates $d {\tilde \omega}_0 /d\tau \equiv d \frac{\omega_0(t)}{\omega_1} / d \frac{t}{t_E}$ between $5\times 10^{-4}$ and $8 \times 10^{-3}$. 
 
We applied the standard CPMG sequence to the rf coil consisting of an initial $90^\circ$ pulse followed by a long train of $180^\circ$ pulses. Standard two-step phase cycling was used. The echo spacing $t_E$ was set between 6.4 $t_{180}$ and 24 $t_{180}$. CPMG trains with up to 8956 echo amplitudes were acquired.
For each measurement $M(t)$ with a time-dependent bias field, an auxiliary measurement $M_{aux}(t)$ without bias field but with otherwise identical parameters was performed. The auxiliary measurement was used to phase both sets of data.  The phased echo amplitudes $M(t)$ were then normalized with respect to the auxiliary measurements to compensate for relaxation effects. We present results in terms of the normalized signal $S(t) \equiv {M(t)}/{M_{aux}(t)}$.
\section{Results}
\label{sec:experiments}

\subsection{Moderate field variations: Adiabatic behavior}

\begin{figure}
	\centering
	\adjincludegraphics[width=3.0in, trim={{.20
			\width} {.05\width} {.20\width} 0},clip]
	{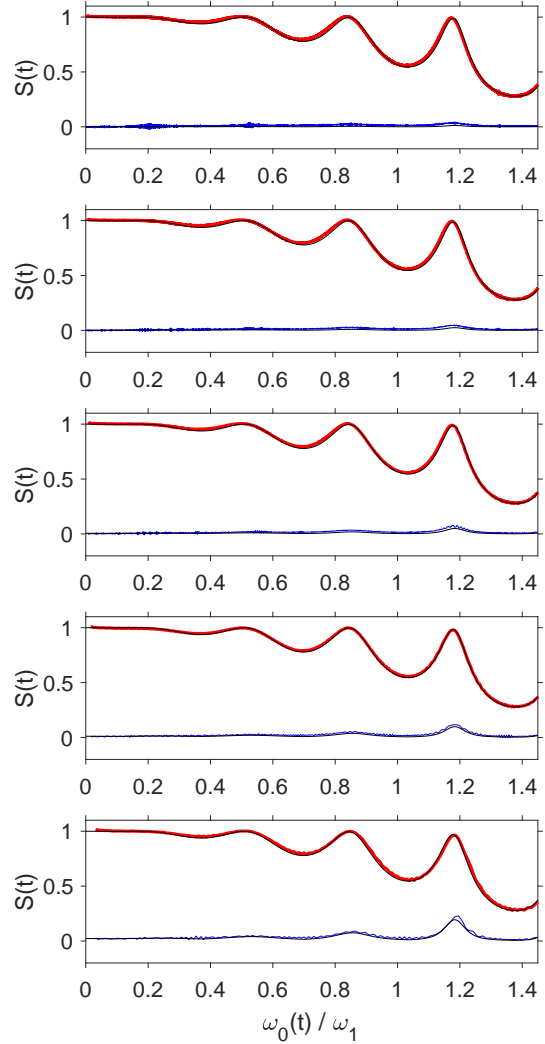}
	\caption[Figure 1]{Normalized in-phase (red) and out-of-phase (blue) echo amplitudes $S(t)$ versus the instantaneous offset in normalized Larmor frequency, $\omega_0/\omega_1$, for a CPMG sequence with a linearly increasing $B_0$ field starting on resonance.  From top to bottom panel, the ramp rates $d {\tilde \omega}_0 /d\tau$ are $5\times 10^{-4}$, $1\times 10^{-3}$, $2\times 10^{-3}$, $4\times 10^{-3}$, and $8\times 10^{-3}$, respectively. The echo spacing was fixed at $t_E = 6.4 t_{180}$. The experimental measurements are well described by the analytical results for adiabatic dynamics given by Eqs.~\ref{eq:Mx-My-first-order}, shown as black lines. }
	\label{fig:Adiabatic_versus_ramprate}
\end{figure}

We first consider the case of a linear field ramp that starts from resonance.
In the experiments shown in Fig.~\ref{fig:Adiabatic_versus_ramprate}, five different normalized ramp rates  $d {\tilde \omega}_0 /d\tau$ between $5\times 10^{-4}$ and $8 \times 10^{-3}$ were used to increase the offset frequency $\omega_0$ from zero (i.e. resonance) to  $1.45 \omega_1$.  
The signal shows systematic modulations as a function of the offset frequency. It is observed to be predominantly in-phase with the refocusing pulses and displays only a weak dependence on the ramp rate. 

These results are consistent with the expected behavior in the adiabatic regime.
The  $90^\circ$ excitation pulse generates transverse magnetization in-phase with the refocusing pulses. In the static on-resonance case, this initial magnetization is an eigenstate of the propagator with unity eigenvalue, i.e. the so-called CPMG subspace. In time-dependent fields, it remains approximately an eigenstate on resonance as long as $\delta \epsilon \equiv \frac{t_{E}^{2}}{8}\frac{d\omega_0}{dt} \ll 1 $. In our experiments,  $\delta \epsilon$ lies between $1.3 \times 10^{-3}$ and $2.0 \times 10^{-2}$ and fulfills this condition well. (For larger echo spacings, the initial magnetization has both CPMG and CP components resulting in the well-known even-odd echo modulation \cite{Carr1954}.)

As the applied field is increased, the initial magnetization remains in the CPMG eigenspace and follows the field variations adiabatically if the ramp rate is smaller than the offset-dependent critical velocity. The  minimum critical velocity in the range of offset frequencies up to $1.45 \omega_1$ is  $\nu_{0,crit} = 0.042$, i.e. more than 5 times  larger than the highest ramp rate $d {\tilde \omega}_0 /d\tau$ used. Therefore, the adiabatic condition ${\cal A} \gg 1$ is fulfilled and the magnetization is effectively spin-locked to the CPMG eigenvector. The observed signal modulation reflects the variable direction of the eigenvector with offset frequency $\omega_0$.  

The CPMG eigenvector of the propagator has no out-of-phase component in the stationary case and its transverse component is given by Eq. \ref{eq:n_perp}. For finite ramp rates, there are slight modification that can be expanded in terms of $1/{\cal A}$. The first order results are given by Eqs.~\ref{eq:Mx-My-first-order} and show some small out-of-phase components proportional to $1/{\cal A}$. For a comparison with the experimental results in Fig.~\ref{fig:Adiabatic_versus_ramprate}, we have convoluted the theoretical expressions in Eqs.~\ref{eq:Mx-My-first-order} with the known relative field inhomogeneity of $6.7\times 10^{-3}$ of the applied offset field,  $\omega_0$.  The experimental results are in excellent agreement with this prediction for full occupation of the CPMG mode, i.e. $a_{CPMG} = 1$. This demonstrates that the modulation of the observed echo magnetization is controlled by the changing direction of the CPMG eigenvector. 

%

\begin{figure}
	\centering
	\adjincludegraphics[width=3.0in, trim={{.20
			\width} {.05\width} {.20\width} 0},clip]%
	{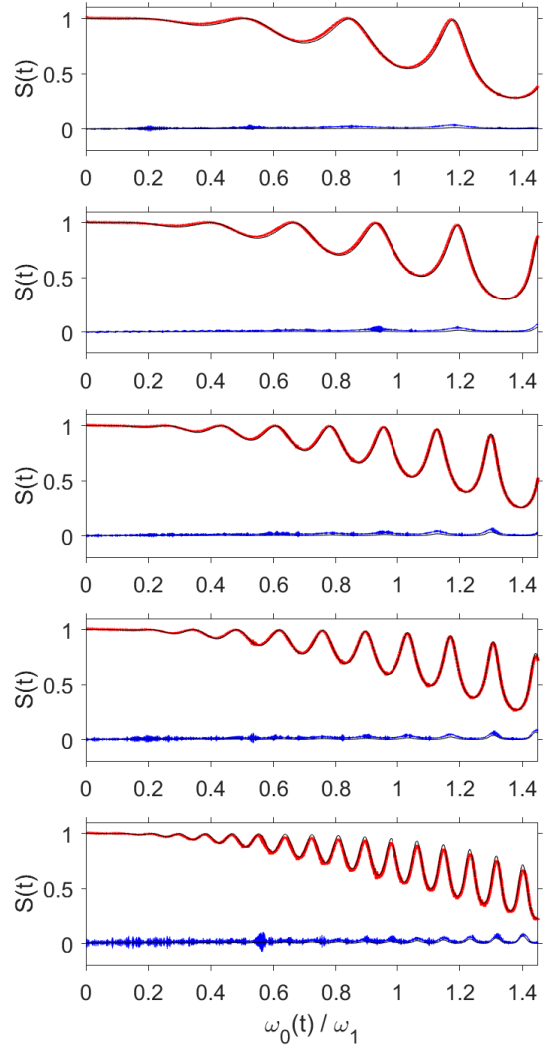}
	\caption[Figure 2]{Measured in-phase (red) and out-of-phase (blue) echo amplitudes for a CPMG sequence with a linearly increasing $B_0$ field, starting on resonance for different echo spacings. From top to bottom panel, the normalized echo spacings $t_E / t_{180}$ are $ 6.4 $, $8.0$, $12.0$, $15.0$, and $24.0$, respectively. The normalized ramp rate $d{\tilde \omega}_0 /d\tau$ was fixed at $5\times 10^{-4}$.  The signal was normalized by standard CPMG signal to account for relaxation. The black lines show the analytical results for adiabatic dynamics given by Eqs.~\ref{eq:Mx-My-first-order}. }
	\label{fig:Adiabatic_versus_tE}
\end{figure}

As a further test, we have measured the response for different echo spacings over the same  range of offset frequencies. Here we kept the normalized ramp rate fixed at $5 \times 10^{-4}$. The results are presented in Fig.~\ref{fig:Adiabatic_versus_tE}.  

The adiabaticity condition ${\cal A} \gg 1$ is again fulfilled for this range of experimental parameters, and we expect adiabatic behavior. The experiments show a more pronounced signal modulation for the measurements with longer echo spacings. This is in quantitative agreement with the expectations for the adiabatic regime for all echo spacings, based on Eq.~\ref{eq:n_perp} and Eqs.~\ref{eq:Mx-My-first-order}.  
%


\subsection{Larger field variations: Observation of non-adiabatic events}
\begin{figure}
	\centering
	\adjincludegraphics[width=5.5in, trim={{.00
			\width} {.20\width} {.00\width} {.25\width}},clip]%
	{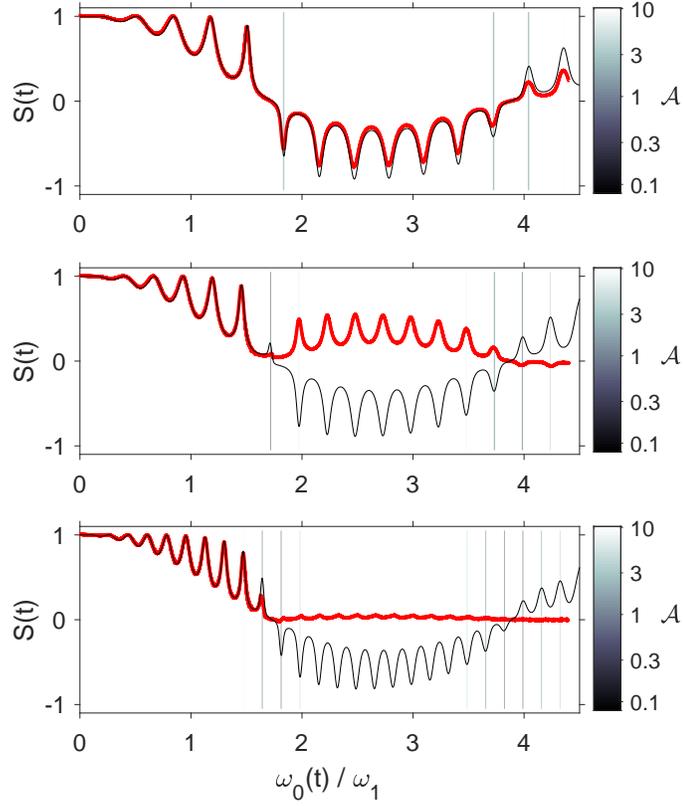}
	\caption[Figure 3]{Measured in-phase (red) 
		echo amplitudes for a CPMG sequence with a linearly increasing $B_0$ field, starting on resonance for different echo spacings. The signal was normalized by standard CPMG signal to account for relaxation. The normalized ramp rate $d{\tilde \omega}_0 /d\tau$ was fixed at $5\times 10^{-4}$. From top to bottom panel, the normalized echo spacings $t_E / t_{180}$ are $ 6.4 $, $8.0$, and $12.0$, respectively. The black lines show the lowest order analytical result, Eq.~\ref{eq:S_fullyadiabatic} for $a_{CPMG}=1$. The calculated values of $\cal A$ are displayed as grey scale. }
	\label{fig:NonAdiabatic_versus_tE}
\end{figure}

When the field ramp extends over a larger range of offset frequencies, deviations from the adiabatic behavior are observed even for  moderate ramp rates. This is apparent from the results presented in  Fig.~\ref{fig:NonAdiabatic_versus_tE}. It shows the in-phase signal for a linear field ramp starting from 0 up to an offset frequency of $4.4 \omega_1$ at a ramp rate of $5\times 10^{-4}$ for three different echo spacings. For small to moderate frequency offsets, the experimental results follow the theoretical expectation for the adiabatic regime as discussed above with $a_{CPMG} = 1$. 

However, beyond a distinct offset frequency, the experimental results abruptly start to deviate from the adiabatic predictions with $a_{CPMG} = 1$. These abrupt changes occur precisely at offset frequencies where the adiabaticity parameter $\cal A$ drops towards 1 or lower, as indicated  in Fig.~\ref{fig:NonAdiabatic_versus_tE} in grey scale. At these offset frequencies, the critical velocity  $\nu_{0,crit}(\omega_0) $ has pronounced minima and the adiabatic condition  ${\cal A} \gg 1$ is not fulfilled anymore. This leads to non-adiabatic transitions between the  CPMG and the CP eigenspace.

During the crossing of these non-adiabatic regions, magnetization is exchanged between the CPMG  and  CP subspaces.  The dispersion of the eigenvalues of the CP mode leads to a quick dephasing of the magnetization associated with the CP subspace in the presence of the slight inhomogeneity in the applied bias field (in analogy to $T_{2}^{*}$ process). The detected signal is therefore dominated by the remaining magnetization of the dispersion-less CPMG mode. After a non-adiabatic region of field offset is passed, the spin dynamics enters again an adiabatic region. The signal follows closely a rescaled version of the adiabatic expression Eq.\,\ref{eq:S_fullyadiabatic}, with a scaling factor $a_{CPMG}$ less than 1 that indicates a reduced occupation of this mode. As is evident from Fig.~\ref{fig:NonAdiabatic_versus_tE}, the amplitudes $a_{CPMG}$ can be either positive or negative.


As the external field is further increased,  additional non-adiabatic regions are encountered that further modify the modal amplitude $a_{CPMG}$. The individual non-adiabatic regions are  well separated and narrow with respect to the normalized offset frequency $\omega_0 / \omega_1$. The results of Fig.~\ref{fig:NonAdiabatic_versus_tE} demonstrate that the adiabaticity parameter ${\cal A}$ is an accurate indicator for the location of the non-adiabatic regions.

It is challenging to predict the change in  amplitude of the CPMG mode $a_{CPMG}$ after passing a non-adiabatic region. The width of these regions is narrow with respect to $\omega_0 / \omega_1$, but generally much wider than the normalized ramp rate. Therefore, the traversing of a particular non-adiabatic region occurs over many refocusing cycles. The net change of the CPMG modal amplitude is the result of the accumulated transition rates between the CPMG and CP subspaces over all these cycles. As was shown in~\cite{Hurlimann2019}, these net rates are sensitive to the experimental parameters and they are affected by the field inhomogeneities. While the location of this non-adiabatic event can be accurately predicted, the associated changes in amplitude are therefore more difficult to predict robustly.

\subsection{Reversibility}
\begin{figure}
	\centering
	\adjincludegraphics[width=4.5in, trim={{.00
			\width} {.08\width} {.00\width} {.05\width}},clip]%
	{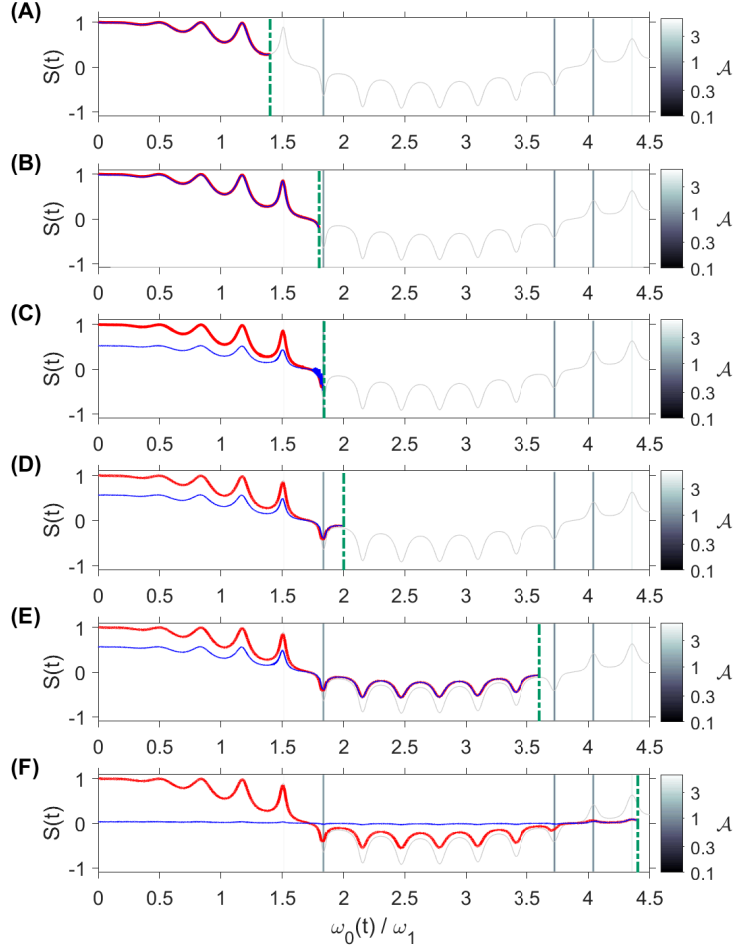}
	\caption[Figure 4]{Measured in-phase  
		echo amplitudes versus instantaneous normalized offset frequency, $\omega_0(t)/\omega_1$ for a CPMG sequence with a bi-linear $B_0$ ramp. The field was linearly ramped up from on-resonance to a value that corresponded to a maximum frequency offset of $\Delta \omega_0$ indicated by the green dashed-dotted lines. The up-ramp was immediately followed by a ramp-down back to on-resonance. The signal for the up-ramp, $S_{\uparrow}$, and the down-ramp, $S_{\downarrow}$, are shown in red and in blue, respectively.   The magnitude of the normalized ramp rates $d{\tilde \omega}_0 /d\tau$ for both up- and down-ramp was fixed at $1\times 10^{-3}$ and the echo spacing was set at $t_E/t_{180} = 6.4$. The light grey curve shows the expected signal for a purely adiabatic behavior, Eq.~\ref{eq:S_fullyadiabatic}, with $a_{CPMG}=1$. The calculated values of $\cal A$ are displayed in grey scale. }
	\label{fig:RTO_versus_offset}
\end{figure}

Further insight can be gained from experiments performed with bi-linear field ramps shown in Fig.~\ref{fig:RTO_versus_offset}. The field was linearly ramped up from on-resonance to a maximum value followed by a ramp-down back to on-resonance.  The magnitude of the ramp rates $\left| d{\tilde \omega}_0 /d\tau \right|$ during the up and down field ramp were fixed at $10^{-3}$. The measurements were repeated with systematically varied maximum field values $\Delta B_0$ that correspond to frequency offsets $\Delta \omega_0 \equiv \gamma \Delta B_0$.  

The results show an interesting  dependence on $\Delta \omega_0 $. When $\Delta \omega_0 $ is smaller than the first non-adiabatic region (panels (A) and (B) in Fig.~\ref{fig:RTO_versus_offset}),  the responses of normalized echo amplitudes versus instantaneous offset frequency during the up-ramp and down-ramp are identical. When the field is returned to its original value, the spin echoes are fully refocused (up to unavoidable relaxation). This reversibility indicates a purely adiabatic behavior with unity eigenvalue for the CPMG mode.  In this regime, the evolution caused by the field fluctuation is fully reversible, including the dephasing caused by the inhomogeneities of the bias field. Note that this is even the case for field variations up to $\Delta \omega_0 = 1.8 \omega_1$ shown in panel (b) in Fig.~\ref{fig:RTO_versus_offset}. During this field sweep, the transverse magnetization temporarily vanishes and becomes negative, all in a fully reversible manner.

However, as soon as $\Delta \omega_0 $ exceeds the critical value of  $1.83 \omega_1$ and the spin dynamics enters the non-adiabatic region, the signal fails to fully refocus when the offset field is returned back to zero. The response during the down-ramp, $S_{\downarrow}(\omega_0)$, is generally rescaled from the response during the up-ramp, $S_{\uparrow}(\omega_0)$. The scaling factor can be interpreted as the amplitude $a_{CPMG}$ of the CPMG mode in Eq.\,\ref{eq:S_fullyadiabatic}. It is a function of the number of non-adiabatic regions encountered during the field sweep. For frequency offsets between the last non-adiabatic region and  $\Delta \omega_0 $, the response is again reversible.

\begin{figure}
	\centering
	\adjincludegraphics[width=5.0in, trim={{.00
			\width} {.17\width} {.00\width} {.17\width}},clip]%
	{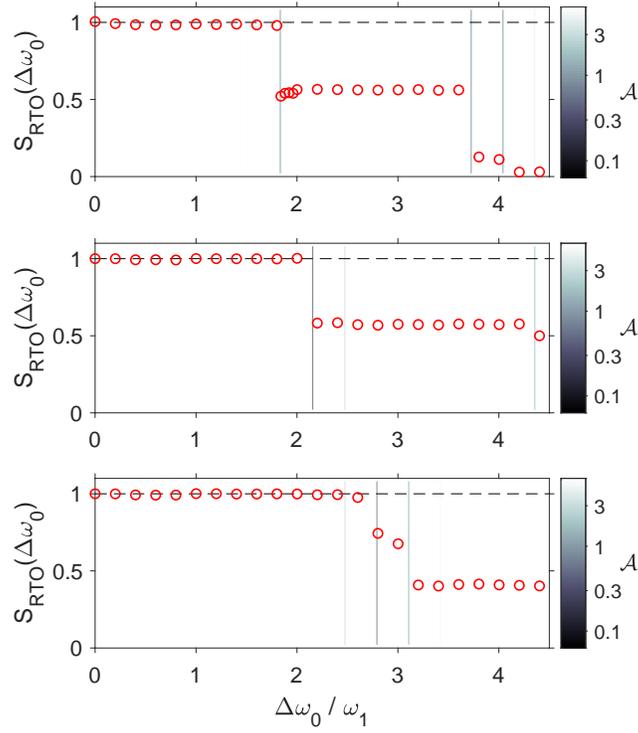}
	\caption[Figure 5]{Measured in-phase signal at the end of the scan, $S_{RTO}$, versus the maximum frequency offset during the sweep,  $\Delta \omega_0$ for the bilinear field ramps shown in Fig.~\ref{fig:RTO_versus_offset}.
	From top to bottom panel, the duration of the refocusing pulses corresponded to a nominal $180^\circ$, $150^\circ$, and $120^\circ$ pulse respectively. The values of $\cal A$ are displayed as grey scale. }
	\label{fig:RTO_versus_HE}
\end{figure}

In Fig.~\ref{fig:RTO_versus_HE}, we plot the signal that is acquired at the end of the scan (when the field has returned to its original value), $S_{RTO}$, versus the maximum frequency offset during the sweep,  $\Delta \omega_0 $. This can be interpreted as the amplitude $a_{CPMG}$ of the CPMG mode at the end of the scan. The results shown in Fig.~\ref{fig:RTO_versus_HE} exhibit a simple step-like function. The locations of the steps coincide exactly with the predicted non-adiabatic regions where $\cal A$  approaches or becomes less than $1$. 

The experimental results of Fig.~\ref{fig:RTO_versus_HE} show a striking simplicity. The degree of refocusing at the end of the scan, $S_{RTO}$, depends only on the number of adiabatic regions that were crossed during the scan. It does not depend on the exact value of $\Delta \omega_0 $ within a particular adiabatic region. These measurements therefore allow a direct identification of the adiabatic and non-adiabatic regions.

Based on numerical simulations with uniform fields reported in \cite{Hurlimann2019}, this simplicity was not anticipated. As the applied field ramps through a non-adiabatic region, some magnetization is converted from the CPMG to CP modes. In the following adiabatic regime, the CP magnetization propagates with non-unity eigenvalues and acquires non-zero dynamical and possibly geometric phases that are sensitive to $\Delta \omega_0 $. After the field is reversed and crosses again the non-adiabatic region, \textcolor{black}{part of the remaining CPMG magnetization will again be converted to the CP mode. In addition,} some of the CP magnetization is converted back to the CPMG mode and is added to the CPMG mode. We expect that the contribution of the CPMG mode converted from the CP mode depends sensitively on $\Delta \omega_0$ due to the variation of the accumulated phase. The absence of a dependence of $S_{RTO}$ on $\Delta \omega_0 $ within an adiabatic region in the experimental results of Fig.~\ref{fig:RTO_versus_HE} indicates that the converted CP magnetization does \textcolor{black}{{not}} make any net contribution to the detected signal $S_{RTO}$.

\textcolor{black}{The absence of sensitivity of $S_{RTO}$ on $\Delta \omega_0$ can be understood in our experiments as follows. As hinted in previous discussions, the process of driving the magnetization (spin) through a non-adiabatic region can be partitioned into three stages within the adiabatic impulse approximation~\cite{Bogdan2005}: (1) the adiabatic evolution before the transition; (2) the non-adiabatic (impulse) transition; (3) the adiabatic evolution after the transition. Similar to the Landau-Zener transition, the transition in the non-adiabatic region can be organized into an unitary matrix describing the transition rates between different states. With the small relative field inhomogeneity, $6.7\times 10^{-3}$, spins subjected to different field strengths essentially follow the same transition matrix in our experiments~\cite{Shevchenko2010}. In particular, because the non-adiabatic transition occurs in a small range of $\omega_0/\omega_1$, no strong phase variations are expected due to the field inhomogeneity in this region. In contrast, the adiabatic evolution of the magnetization for CP mode are strongly affected by the inhomogeneity of the applied bias field, c.f.~\ref{app:C}, that, in general, induce a phase spreading of the CP components for spins subjected to different field strengths and hence leads to the dephasing, \textit{i.e.}, shorter $T_2^*$, for the CP components. As shown in \ref{app:C}, except at some specific offset frequency, $\omega_0$, where the phase spreading of CP modes is reverted due to the dispersion of the CP mode phase spectrum, the magnetization of CP modes is effectively diminished due to this phase spreading even for the small relative field inhomogeneity in our experimental setup. In addition to the field inhomogeneity effect, the CP components are also preferentially attenuated by diffusion effects.}
	
\textcolor{black}{Now, when an ensemble of CP modes with randomized phases is driven through the non-adiabatic region, the CP modes converted back to CPMG modes will still retain these randomized phases. As a result, the ensemble-averaged CPMG signals converted from these CP modes becomes negligible, which leads to the $\Delta \omega_0 $ independent $S_{RTO}$ signal. This occurs as long as the accumulated phases of these CP modes are well-randomized before entering the non-adiabatic region. On the other hand, part of the CPMG modes will still be converted to CP modes, which will be quickly dephase due to the field inhomogeneity. This implies that the magnitude of $S_{RTO}(\Delta \omega_0)$ should decrease monotonically with every additional non-adiabatic event, as is indeed observed in our experiments.}

\subsection{Measurements with modified refocusing cycles}

The response in time-dependent fields can be modified by replacing the standard $180^\circ$ refocusing pulses with composite or phase-modulated pulses \cite{hurlimann_composite,borneman2010}. It is possible to make the system less or more sensitive towards field fluctuations. Such optimization can be considered a form of Hamiltonian engineering. The current analysis shows that the key quantity to optimize is the critical velocity $\nu_{0,crit}(\omega_0)$, which in turn is determined by the modal properties of a single refocusing cycle. 

To find sequences more robust towards field fluctuations, optimal control based algorithms \cite{khaneja2005,borneman2010,Mandal2014_SNR} can be used to systematically search for refocusing cycles with modal structures that have no near degeneracies and associated low critical velocites for the range of offset frequencies of interest. This will eliminate non-adiabatic regions and result in a fully adiabatic spin dynamics. It could also be desirable to minimize the signal modulation with variable offset frequency. This could be accomplished by searching for refocusing cycles that maximize the critical velocities  while simultaneously minmize the variability in the direction of the eigenvector of the CPMG mode versus $\omega_0$.   

\begin{figure}
	\centering
	\adjincludegraphics[width=5.5in, trim={{.00
			\width} {.20\width} {.00\width} {.25\width}},clip]%
	{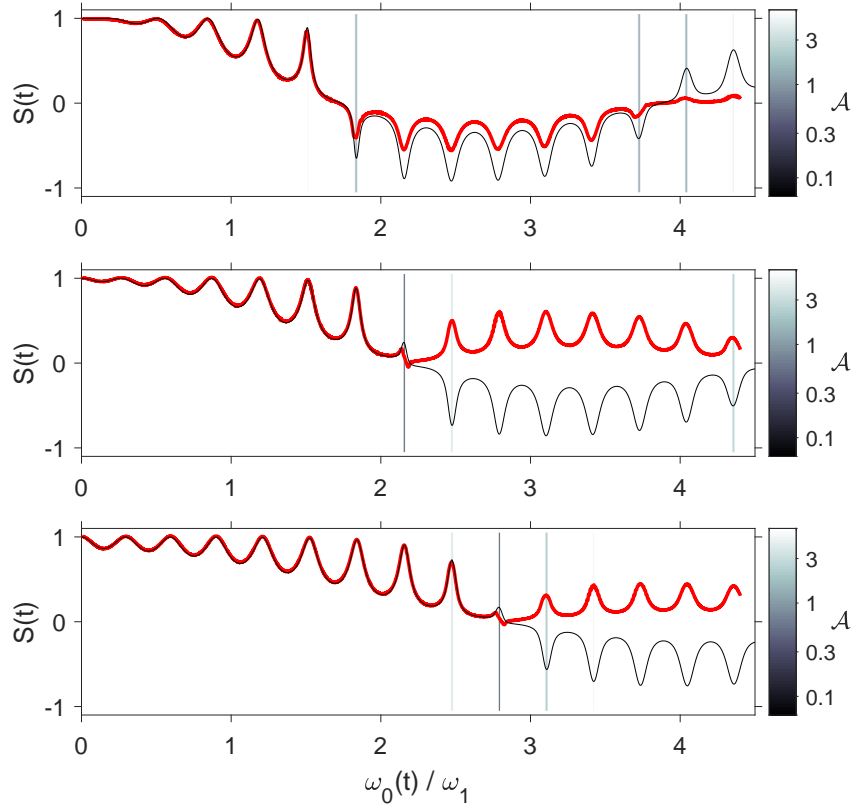}
	\caption[Figure 6]{Measured in-phase (red) 
		echo amplitudes for a CPMG sequence with a linearly increasing $B_0$ field, starting on resonance for different refocusing pulses. From top to bottom panel, the duration of the refocusing pulse corresponded to a nominal $180^\circ$, $150^\circ$, and $120^\circ$ pulse respectively.  The normalized ramp rate $d{\tilde \omega}_0 /d\tau$ was fixed at $1\times 10^{-3}$ and the echo spacing at $t_E/t_{180} = 6.4$. The signal was normalized by standard CPMG signal to account for relaxation. The black lines show the theoretical expectation for the fully adiabatic case based on Eq.~\ref{eq:S_fullyadiabatic}. The values of $\cal A$ are displayed as grey scale. }
	\label{fig:NonAdiabatic_versus_HE}
\end{figure}

As a simple illustration of this general approach of modified CPMG sequences, we present in Fig.\,\ref{fig:NonAdiabatic_versus_HE} measurements with shortened refocusing pulses. The pulse durations were reduced from the default value of $t_{180} = \pi/\omega_1$ by $16 \%$ or $33 \%$  without increasing the pulse amplitude. The pulses now correspond to nominal $150^\circ$ or $120^\circ$ pulses, respectively. The analytical expressions of the modal properties of the propagator in \ref{app:A} show that a reduction of the pulse durations moves the location of the non-adiabatic regions to higher offset frequencies, thus increasing the range of the first adiabatic region. This prediction is confirmed by the experimental measurements of Fig.\,\ref{fig:NonAdiabatic_versus_HE}: the echo amplitudes with the shortened refocusing pulses follow the adiabatic result to a significantly larger offset frequency. The modified sequences have an improved robustness towards the amplitude of field fluctuations that can be fully refocused. As shown in Fig.\,\ref{fig:RTO_versus_HE}, with $120^\circ$ refocusing pulses field fluctuations up to about 2.8 $B_1$ can be fully recovered without loss of signal, while with standard $180^\circ$ refocusing pulses this can only be achieved with fluctuations up to about 1.8 $B_1$. However, this increased adiabatic range is associated with a more pronounced modulation of the echo amplitudes at small offset frequencies. It might be possible to find more complex refocusing pulses that can eliminate this drawback of enhanced signal modulation but retain the increased adiabatic range.

\section{Conclusion}
\label{sec:conclusions}

We have applied and tested a new framework to characterize the response of CPMG-like sequences to time-dependent magnetic fields. The approach is based on the decomposition of the magnetization in terms of the eigenmodes of the propagator. We have shown that the response can be generally classified into adiabatic behavior that is occasionally interrupted by non-adiabatic events.

In the adiabatic regime, the simple analytical expressions have been experimentally confirmed. In the presence of moderate field inhomogeneities, the magnetization associated with the CP modes generally dephases quickly. The detected magnetization is then dominated by the magnetization of the CPMG eigenmode that is effectively spin-locked to its eigenvector and robust towards field inhomogeneities. Changes in the occupation of the eigenmodes requires a non-adiabatic event. They occur in narrow ranges of offset frequencies where the adiabaticity condition ${\cal A} \gg 1$ is not fulfilled. The adiabaticity parameter  ${\cal A}$ is the ratio of an intrinsic, offset dependent critical velocity that is derivable from modal properties of the propagator, and of the instantaneous ramp rate of the applied field.

The approach applies more generally to any CPMG-like sequence that consists of an initial excitation pulse followed by  identical refocusing cycles. The refocusing cycle can not only consist of a single RF pulse, but can contain any composite pulse, a combination of RF pulses, or frequency sweeps. Thus, it is possible to optimize the response to time-dependent magnetic fields by finding refocusing cycles with the desired properties of the critical velocity and other modal properties. Sequences that are more robust towards temporal field variations are associated with high critical velocities without any pronounced minima in the relevant range of offset frequencies. Alternatively, it might be desirable to find sequences that show an enhanced sensitivity to field fluctuations. In that case, the goal is to use an operating point with a low critical velocity, near a singular point where the different modes become degenerate and modal transitions occur even with small field fluctuations.   

\textcolor{black}{From the reversibility test with the bi-linear field ramps to a maximum offset frequency, $\Delta \omega_0$, we observe a rather simple step-like structure of the returning CPMG signals with respect to the number of the encountered non-adiabatic events. We argue that this is due to the absence of CP mode contributions in an inhomogeneous applied field environment. This surprisingly simple response indicates the robustness for the measured CPMG magnetization. On the other hand, it also hinders the observation of the interference effects between two or multiple non-adiabatic transition events~\cite{Hurlimann2019}. By carefully inspecting the in- and out-of phase signals during the ramping process, we observe the refocusing of the CP mode signals at specific de-tuning frequencies, c.f.~\ref{app:C}, with the mechanism similar to that for the gradient echo. Hence, with the reduction of the field inhomogeneity, we are hopeful that the interference effects can be observed. Coupled with Hamiltonian engineering through the composite pulse sequence, it would be interesting to explore the potential of using our system as simulator to probe dynamical responses and interference phenomena for certain quantum systems~\cite{Shevchenko2010,Gustavsson2013,Ferron2017}.
}     
 
\appendix

\section{Properties of eigenmodes of refocusing cycle with a rectangular RF pulse}
\label{app:A} 

We consider a refocusing cycle of duration $t_E$ that contains a single rf pulse of duration $t_p$ centered in the middle of the cycle. The rf pulse is linearly polarized with a carrier frequency of $\omega_{rf}$ and an amplitude $B_{1,\perp}$, resulting in a nominal nutation frequency of $\omega_1 \equiv \gamma B_{1,\perp}/2$. The static field is $B_0$, resulting in an offset between Larmor frequency and applied rf frequency of $\omega_0(t) \equiv \gamma B_0(t) - \omega_{rf}$. For independent spins 1/2, the echo-to-echo evolution can then be described using average Hamiltonian theory by an effective magnetic field $\gamma {\vec B}_{eff} = \alpha {\hat n}$, where $\gamma$ is the gyromagnetic ratio. For a rectangular rf pulse, the direction $\hat n$ and amplitude $\alpha$ of this resulting effective magnetic field can be obtained from the following expressions:
\begin{eqnarray}
n_{\perp} & = & \frac{1}{\Delta}\frac{\omega_1}{\Omega} {\sin{\beta_2}}
\label{eq:n_perp} \\
n_z       & = & \frac{1}{\Delta}\left[{\sin{\beta_1}\cos{\beta_2}+\frac{\omega_0}{\Omega}\cos{\beta_1} \sin{\beta_2}}\right]
\label{eq:n_z} \\
\cos\left( \frac{\alpha}{2}\right) & = & \cos{\beta_1}\cos{\beta_2}-\frac{\omega_0}{\Omega}\sin{\beta_1} \sin{\beta_2},
\label{eq:alpha}
\end{eqnarray}
where
\begin{eqnarray}
\Omega   & = & \sqrt{\omega_{0}^{2}+ \omega_{1}^{2}}\\
\beta_1  & = &\omega_0 (t_E - t_{p})/2 \\
\beta_2  & = &\Omega t_{p}/2 \\
\Delta^2 & = &\left[ \frac{\omega_1}{\Omega}\sin{\beta_2}\right]^2 + \left[ \sin{\beta_1} \cos{\beta_2}+\frac{\omega_0}{\Omega}\cos{\beta_1}\sin{\beta_2}\right]^2.
\end{eqnarray} 

\section{Analytical results for adiabatic regime}
\label{app:B}

In the presence of field inhomogeneities, the detected signal is dominated by the contribution of the CPMG mode. In the fully analytical limit of vanishing small ramp rates $d {\tilde \omega}_0 /d\tau$, the magnetization of the CPMG mode is spin-locked to the eigenvector $\hat n$ derived from the static case. To lowest order in the ramp rate, the signal is given by:
\begin{equation}
S(t) = a_{CPMG}\, n_{\perp}(t).
\label{eq:S_fullyadiabatic}
\end{equation}  
Here $a_{CPMG}$ is the amplitude or occupation of the CPMG mode and $n_{\perp}(t)$ is the transverse component of $\hat n$, \eqref{eq:n_perp}, at the instantaneous offset frequency $\omega_0 (t)$. In this limit, the signal is purely in-phase with the refocusing pulses.

For finite ramp rates, the eigenvector for the CPMG eigenmode acquires a small out-of-phase component. A perturbation calculation to first order in $1/\mathcal{A}$ yields:
\begin{equation}
\label{eq:Mx-My-first-order}
\begin{split}
S_x^{(1)}(t) & = \frac{a_{CPMG}}{\sqrt{1+ 1/\mathcal{A}(t)^2}} \left[ \cos (\delta \epsilon) n_{\perp}(t) - 1/\mathcal{A}(t) \sin (\delta \epsilon) \right],
\\
|S_y^{(1)}(t)| & = \frac{a_{CPMG}}{\sqrt{1+ 1/\mathcal{A}(t)^2}} \left[ \sin (\delta \epsilon) n_{\perp}(t) +  1/\mathcal{A}(t) \cos (\delta \epsilon)  \right].
\end{split}
\end{equation}
Here $\delta \epsilon \equiv \frac{t_{E}^{2}}{8}\frac{d\omega_0}{dt}$.

\section{Effect of field inhomogeneities on signal generated by the CP modes.}
\label{app:C}

In the continuous limit, the dynamic phase of the CP mode in the adiabatic regime for a linear field ramp is given by:
\begin{equation}
\phi_{CP}({\tilde \omega}_0) = \left( \frac{d{\tilde \omega}_{0}}{d\tau} \right)^{-1}
\int_{{\tilde \omega}_{0,start}}^{{\tilde \omega}_0} \alpha({\tilde \omega}_{0}^{'}) d{\tilde \omega}_{0}^{'}.
\label{eq:phi_CP}
\end{equation}
Here ${{\tilde \omega}_{0,start}}$ is the normalized offset frequency where the CP mode has been initialized.
When the applied field $\omega_0$ is non-uniform across the sample, the phase of the CP component is also non-uniform. This leads to a $T_2^*$ - like decay of the detected signal of the CP component, $S_{CP}$. Assuming that the inhomogeneities of the applied field is characterized by a Gaussian distribution with a standard deviation $\sigma_{{ \omega}_0}$, the resulting distribution of the phases has a standard deviation that is to first order given by 
\begin{equation}
\sigma_{\phi_{CP}}^2 = \sigma_{{ \omega}_0}^2 \left( \frac{d{\tilde \omega}_0}{d\tau} \right)^{-2}
\left[\int_{{\tilde \omega}_{0,start}}^{{\tilde \omega}_0} {\tilde \omega}_{0}^{'} \frac{d\alpha({\tilde \omega}_{0}^{'} )}{d {\tilde \omega}_{0}^{'}} d{\tilde \omega}_{0}^{'} \right]^2.
\label{eq:sigma_CP}
\end{equation}  
The field inhomogeneity reduces the detected CP signal by $\exp\left\{-\sigma_{\phi_{CP}}^{2}/2\right\}$ compared to the case of uniform $\omega_0$. In the experimental results presented above, the condition $\sigma_{\phi_{CP}}^{2}({\tilde \omega}_0) \gg 1 $ is generally well fulfilled in the adiabatic regime except in the near vicinity of the non-adiabatic regimes. Consequently, the detected signals are dominated by the CPMG modes and the CP modes make no significant contributions. 

However, there are conditions when $\sigma_{\phi_{CP}}^{2}$ becomes small or even vanishes, even in the presence of field inhomogeneities. In such cases, signals from the CP modes are detected. 
This is made possible by the oscillatory nature of ${d\alpha}/{d {\tilde \omega}_0} $ with $\tilde{\omega}_0$. At particular offset frequencies, the integral in Eq.\,\ref{eq:sigma_CP} becomes zero. At these special points, the overall phase of the CP contribution is to first order independent of the field inhomogeneity (but in general non-zero) and the CP modes form gradient-echo like signals.  Given the $\left( \frac{d{\tilde \omega}_0}{d\tau} \right)^{-2}$ dependence of $\sigma_{\phi_{CP}}^{2}$, such gradient echoes are more pronounced at higher ramp rates.

\begin{figure}
	   \centering
      \adjincludegraphics[width=5.5in, trim={{.00
   		\width} {.29\width} {.00\width} {.30\width}},clip]
        {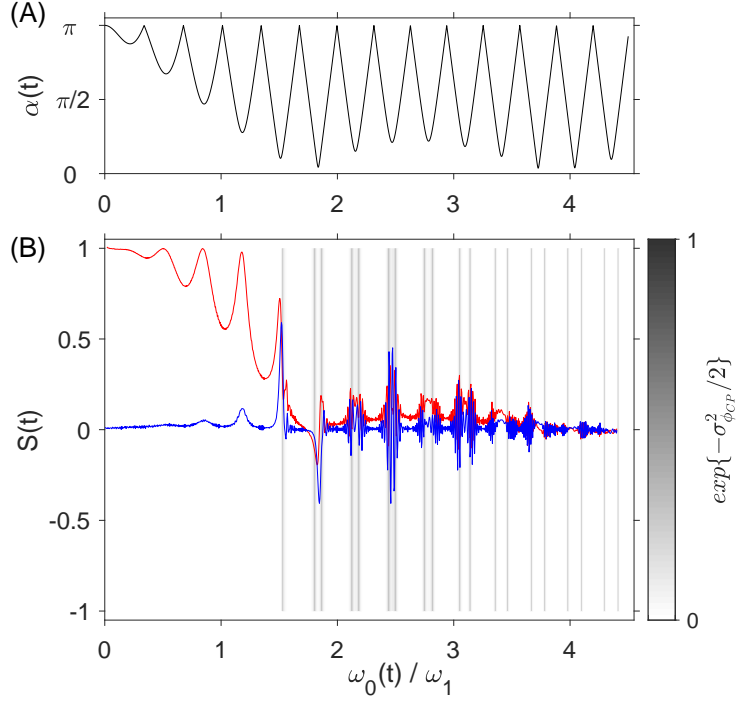}
  	 \caption[Figure 7]{Experimental demonstration of gradient-echo like signals generated by the CP contributions. (A) shows the oscillatory nature of the instantaneous values of $\alpha$ versus the normalized offset frequency ${\tilde \omega}_0(t) = \omega_0(t)/\omega_1$, calculated from Eq.\,\ref{eq:alpha}. In (B), the red and blue lines show the experimental in-phase and out-of-phase signal detected during a linear ramp versus  ${\tilde \omega}_0(t)$ with $d{\tilde \omega}_0 / d\tau =  4\times 10^{-3}$ and $t_E/t_{180} = 6.4$. The grey scale shows the simple prediction of $\exp\left\{-\sigma_{\phi_{CP}}^{2}/2\right\}$. This can be interpreted as the visibility of gradient echoes, as discussed in the text. }
	   \label{fig:gradient_echoes}
\end{figure}

An example of such CP gradient echoes are shown in the experimental results of Fig.\,\ref{fig:gradient_echoes}. This data was acquired at a ramp rate $d{\tilde \omega}_0 / d\tau =  4\times 10^{-3}$ and $t_E/t_{180} = 6.4$. At distinct offset frequencies, oscillating out-of-phase signals show the formation of CP gradient echoes. The position of these echoes corresponds well to the locations where $\exp\left\{-\sigma_{\phi_{CP}}^{2}/2\right\}$ approaches 1. Here $ \sigma_{\phi_{CP}}$ was calculated from Eq.\,\ref{eq:sigma_CP} using ${\tilde \omega}_{0,start}$ as the end of the first non-adiabatic regime encountered. This calculation is a simplification. It implicitly assumes that only the first non-adiabatic event generates CP contributions and that the phase shifts induced by the subsequent non-adiabatic regions can be ignored. Despite these approximation, it gives a good qualitative indication where CP gradient-echoes can form. 
 
\newpage
\bibliographystyle{elsarticle-num}
\bibliography{NMR_MDH_2}

\end{document}